\documentclass[showpacs,twocolumn,APS]{revtex4}

\usepackage{graphicx}
\usepackage{dcolumn}
\usepackage{amsmath}
\makeatletter
\def\btt#1{\texttt{\@backslashchar#1}}
\DeclareRobustCommand\bblash{\btt{\@backslashchar}} \makeatother

\begin{document}
\title[]{Universal features of gravity and higher dimensions}

\author{Naresh Dadhich {\footnote{e-mail address:{nkd@iucaa.ernet.in}}}}
\affiliation{Inter-University Centre for Astronomy \& Astrophysics,\\ Post Bag 4 Pune 411 007, India}
\date{\today}

\begin{abstract}
We study some universal features of gravity in higher dimensions and by universal we mean a feature that remains true in all dimensions $\geq4$. They include:  (a) the gravitational dynamics always follows from the Bianchi derivative of a homogeneous polynomial in Riemann curvature and it thereby characterizes the Lovelock polynomial action, (b) all the $\Lambda$-vacuum solutions of the Einstein-Lovelock as well as pure Lovelock equation have the same asymptotic limit agreeing with the $d$ dimensional Einstein solution and (c) gravity inside a uniform density sphere is independent of the spacetime dimension and it is always given by the Schwarzschild interior solution.
\end{abstract}


\maketitle
\section{Introduction}

In this paper we shall discuss three universal features of gravity in higher dimensions. By universal we mean a feature that is carried over to all dimensions $\geq4$. Gravity is the universal force and hence it could only be described by the curvature of spacetime \cite{d1}. That is the gravitational equation of motion should follow from the Riemann curvature. It does indeed do so because the trace of the vanishing Bianchi derivative of the Riemann curvature (Bianchi identity) yields a second order differential operator in terms of the divergence free second rank symmetric tensor, the Einstein tensor. This should always happen even when we include higher order terms in Riemann curvature. There would always exist an analogue the Einstein tensor for the analogue of Riemann defined as a homogeneous polynomial in Riemann. This is the first universal feature we shall establish by obtaining the gravitational equation from the vanishing of the trace of the Bianchi derivative of the homogeneous polynomial tensor which will characterize the Lovelock polynomial action  \cite{bia}.\\

Higher order terms in the Lovelock action should make significant contribution only on the high energy end, $r\to0$ in the vacuum solution while the low energy asymptotic limit, $r\to\infty$ should always tend to the Einstein solution in the corresponding dimension. We shall show that all pure Lovelock as well as Einstein-Lovelock $\Lambda$ vacuum solutions asymptotically tend to the corresponding Einstein solution \cite{vac}. This is the second universal feature. It is well known that gravitational field inside a uniform density sphere is proportional to $r$ in Newtonian gravity as well as in Einstein gravity despite its non-linearity. This is a general feature that should remain true in all dimensions as well as in higher order Einstein-Lovelock gravity \cite{unif}. This means it would be always described by the Schwarzschild interior solution. This is the third and final universal feature we would establish. \\

The paper is organised as follows: Before taking up the main theme of the paper we would also like to allude on some of the classical motivations for higher dimensions. That is what would be done in the next section. Then we have three sections devoted to the three  universal features which would be followed to conclude in the discussion.

\section{Why higher dimsneions?} 

We shall consider two very general classical motivations for higher dimensions. Since gravity is described by the Riemann curvature of spacetime and hence gravitational equation of motion would remain valid in all dimensions where Riemann curvature is defined. For this the minimum dimension required is $2$ and hence $d\geq2$. However it turns out that $2$ and $3$ dimensions are not big enough to accommodate free propagation and hence we land in the usual $4$ dimension. Now the question arises, are $4$ diemsnions sufficient for description of gravity? Is there, like the free propagation, any other property that still remains untapped? How about the high energy effects - how do we extend a theory to include high energy modification? The usual method for that is to include the non-linear terms in the basic variable. For gravity, the Riemann curvature is the basic variable and so we should include its square and higher orders in the action. However the resulting equation should still remain second order quasi-linear (second derivative must occur linearly) so that the equation describes a unique evolution for a given initial value formulation. It turns out that this can only happen for a particular combination, known as the Lovelock polynomial \cite{vac}. However the higher order Lovelock terms make non-zero contribution in the equation of motion only for $d\geq4$. Thus we have to go to higher dimensions to realize high energy effects \cite{gb}.

Next we appeal to the general principle that the total charge for a classical field must always be zero globally \cite{gb}. This is true for the  electromagnetic field, how about gravity? The charge for gravity is the energy momentum that is always positive, how could that be neutralized? The only way it could be neutralized is that the gravitational field the matter creates itself should have charge of opposite polarity. That means the gravitational interaction energy must be negative and that is why the field has always to be attractive. Note that gravity is attractive to make the total charge zero. The negative charge is however not localizable as it is spread all over the space. When it is integrated over the whole space for a mass point, it would perfectly balance the positive mass.  This is what was rigorously established in the famous ADM calculation \cite{adm}. Consider a neighbourhood of radius $R$ around a mass point. In this region, the total charge is not zero and there is overdominance of positive charge as the negative charge lying in the field outside $R$ has been cut off. Whenever the charge is non zero on any surface (like for an electrically charged sphere, the field propagates off the sphere) the field must go off it. This means gravity must go off the $3$-brane ($4$-spacetime) in the extra dimension but as it leaks out, its past light cone would encompass more and more region outside $R$ and thereby more and  more the negative charge. That is it leaks off the brane with diminishing field  strength and hence it does not penetrate deep enough. This is the picture quite similar to the Randall-Sundrum braneworld gravity model \cite{rs}. If the matter fields remain confined to the $3$-brane and only gravity leaks into extra dimensions but not deep enough, then extra dimensions effectively become small (for probing depth of dimension, we need some physical probe that goes there). This is an intuitively very appealing and enlightening classical consideration why gravity cannot remain confined to $4$ dimension and at the same time why extra dimension cannot be large?

\section{Bianchi derivative}

Gravity is the universal force in the sense that it links to everything including the zero mass particles which demands that it can only be described by the curvature of spacetime.  This means its  dynamics has always to be goverened by the spacetime geometry - a property of the curvature. That is, the gravitational field equation should follow from the Riemann curvature. The Riemann curvature satisfies the Bianchi identity which on contraction yields the second rank symmetric Einstein tensor with vanishing divergence. This provides the appropriate second order differential operator required for deducing the equation of motion for gravitation - the Einstein equation. This should always happen even when higher order terms in curvature are included. It turns out that it is possible to identify the analogue of Riemann tensor which is a homogeneous polynomial in Riemann, and it is the trace of the Bianchi derivative of this polynomial would provide the analogue of the Einstein tensor \cite{bia}. This leads to a geometric characterization of the Lovelock polynomial. \\

In here, we would refer to high energy behavior of gravity, that would ask for inclusion of higher order terms in the curvature. Usually such corrections are evaluated as perturbations against a fixed background spacetime provided by the low energy solution. Since at high energy, spacetime curvature would also be very strong, hence we cannot resort to the usual perturbative analysis but instead have to consider the situation non-perturbatively. That is, include higher order terms in Riemann curvature, derive the equation of motion and then seek its solution. For the next order, we include square of the Riemann curvature and yet we must have a second order quasilinear (linear in the second derivative) equation for unique evolution with a proper initial data. This uniquely identifies a particular combination, $L_{GB} = R_{abcd}^2 - 4R_{ab}^2 + R^2$, (where $R_{ab}^2 = R_{ab}R^{ab}$) known as the Gauss-Bonnet (GB) Lagrangian. This has remarkable property that the squares of the second derivative get canceled out leaving the equation quasilinear. The variation of this 
as well as the trace of the Bianchi derivative of a fourth rank tensor which is a homogeneous quadratic in Riemann curvature leads to the analogue of $G_{ab}$, a divergence free $H_{ab}$. \\

We define an analogue of $R_{abcd}$ as a homogeneous polynomial \cite{bia} as 
\begin{equation}
F_{abcd} = P_{abcd} - \frac{n-1}{n(d-1)(d-2)} P(g_{ac}g_{bd} - g_{ad}g_{bc}) \label{F}
\end{equation}
where 
\begin{equation}
P_{abcd} = Q_{ab}{}^{mn}R_{mncd}
\end{equation}
and 
\begin{equation}
Q^{ab}{}_{cd} =
 \delta^{aba_1b_1...a_nb_n}_{cdc_1d_1...c_nd_n}R_{a_1b_1}{}^{c_1d_1}...R_{a_nb_n}{}^{c_nd_n}.
\end{equation}
Here $n$ is the order of polynomial and $Q^{abcd}{}{}{}{}_{;d}=0$. For the quadratic case, $P_{abcd}$ reads as 
\begin{equation}  
P_{abcd} = R_{abmn}R_{cd}{}^{mn} + 4R_[{a}{}^{m}R_{b]mcd} + R R_{abcd} 
\end{equation}
where $P = L_{GB}$. Note that the Bianchi derivative of $F_{abcd}$ does not vanish (that only vanishes for $R_{abcd}$), however its trace does vanish to give 
\begin{equation}
 -\frac{n}{2}  F^{cd}{}{}_{[cd;e]} = H_e{}^c_{;c} = 0
\end{equation} 
where 
\begin{equation}
n( F_{ab} -\frac{1}{2} F g_{ab}) = H_{ab}.
\end{equation}
This is an alternative derivation of $H_{ab}$ which results from the variation of the corresponding $n$-th order term in the Lovelock polynomial which is defined by $Q_{abcd}R^{abcd}$. This is yet another characterization, the trace of the Bianchi derivative yielding the correponding $H_{ab}$, of the Lovelock gravity. In the GB quadratic case by varying the Lagrangian $L_{GB} = R_{abcd}^2 - 4R_{ab}^2 + R^2$ we obtain 
\begin{eqnarray}
H_{ab}& =& 2(RR_{ab} - 2R_{a}{}^{m}R_{bm} - 2R^{mn}R_{ambn}\nonumber \\  
&+& R_{a}{}^{mnl}R_{bmnl}) - \frac{1}{2} L_{GB} g_{ab}.
\end{eqnarray}
However $H_{ab}$ is non-zero only for $d>4$ which means GB and higher order Lovelock make  non-zero contribution in the equation of motion only in dimension higher than four. This clearly indicates that at high energies gravity cannot remain confined to the four dimensions and hence the consideration of higher dimensions becomes pertinent and relevant. Its dynamics in higher dimensions would be governed by $H_{ab} = -\Lambda g_{ab}$. \\

Since the corresponding $H_{ab}$ could always be gotten from the trace of the Bianchi derivative for any Lovelock order, gravitational dynamics would thus entirely follow from the geometric properties of spacetime curvature. This is a universal feature which is true for all Lovelock orders and in all dimensions $\geq4$.\\

\section{Lovelock vacuum solutions}


Although the Lovelock vacuum solutions are known for long time \cite{whitt}, \cite{whee},  \cite{mey} we would here like to probe the universality of their asymptotic large $r$ behaviour \cite{vac}. It turns out that this limit is neutral for pure Lovelock solution and its Einstein-Lovelock analogue so long as $\Lambda$ is non-zero. In particular, the asymptotic limit of the Einstein-Gauss-Bonnet solution with $\Lambda$ has the same form as the pure Gauss-Bonnet solution and it is the $d$-dimensional Einstein solution. However their $r\to 0$ limit is radically  different, for the former the metric remains regular and finite \cite{gb} while for the latter it is singular at $r=0$. We would also like to draw attention to an interesting property of spherically symmetric vacuum equations that one has ultimately to solve a single first order equation not withstanding the enhanced nonlinearity of the Lovelock gravity. \\

We shall begin with the GB vacuum equation,
\begin{equation}
H_{ab} = -\Lambda g_{ab}
\end{equation} 
where $H_{ab}$ as given in Eq (7), for the spherically symmetric metric, 
\begin{equation}
ds^2= e^\nu dt^2 - e^\lambda dr^2 - r^2d\Omega_{d-2}^2
\end{equation}
where $d\Omega_{d-2}^2$ is the metric on a unit $(d-2)$-sphere. In general $\nu, \lambda$ are functions of both $t$ and $r$, however as shown in \cite{whitt} the $t$-dependence drops out as usual and it then suffices to take them as functions of $r$ alone. To begin with we have $H^t_t = H^r_r$ that immediately determines $\nu = -\lambda$. With this, let us write the non-zero components of $H^a_b$ for the above metric and they read as follows: 
\begin{equation} 
H^t_t = H^r_r = -\frac{d-2}{2r^4}(1-e^{-\lambda}) \bigl(2re^{-\lambda}\lambda^{\prime} + (d-5)(1-e^{-\lambda}) \bigr)  
=-\Lambda 
\end{equation}
\begin{eqnarray}
H^\theta_\theta&=& \frac{1}{2r^4} \Bigl[r^2e^{-\lambda}(1-e^{-\lambda})(-2\lambda^{\prime \prime} + \lambda^{\prime^2})  
- r^2e^{-2\lambda}(3-e^{\lambda})\lambda^{\prime^2} \nonumber \\
&+& (d-5)(1-e^{-\lambda})\bigl(-4re^{-\lambda}\lambda^{\prime} - (d-6)(1-e^{-\lambda}) \bigr) \Bigr] \nonumber \\
&=& -\Lambda 
\end{eqnarray}
where a prime denotes derivative w.r.t $r$ and all the angular components are equal. \\

First let us note that the above two equations are not independent and it can easily be seen that the latter is a derivative of the former which was first shown for the usual four dimensional gravity in \cite{d-sch}. It would therefore suffice to integrate the former
 alone to get the general solution. This is what it should be because there is only one function, $\lambda$, to be determined. Eq (10) could be written as 
\begin{equation}
(r^{d-5}f^2)^\prime = \frac{2\Lambda}{d-2}r^{d-2}
\end{equation}
which readily integrates to give 
\begin{equation} 
e^{-\lambda} = F = 1-f, ~~ f^2 = \frac{k}{r^{d-5}} + \Lambda_1 r^4
\end{equation}
where $2\Lambda/{(d-1)(d-2)} = \Lambda_1$. This is the general solution of the pure Gauss-Bonnet vacuum which has been obtained by solving the single first order equation. Its large $r$ limit would be
\begin{equation} 
F =1-\sqrt{\Lambda_1}r^2-\frac{K}{r^{d-3}} 
\end{equation}
where $K = k/{2\sqrt{\Lambda_1}}$. This is the Schwarzschild-dS solution for a $d$-dimensional spacetime. On the other hand let us look at the Einstein-Gauss-Bonnet solution (which is obtained by summing over $n=1,2$;i.e. replacing $H_{ab}$ in Eq (8) by $G_{ab} + \alpha H_{ab}$) \cite{bd}, 
\begin{equation} 
F = 1+\frac{r^2}{2\alpha}[1-\sqrt{1+4\alpha(\frac{M}{r^{d-1}}+\Lambda)}] 
\end{equation}
that would also approximate for large $r$ to 
\begin{equation} 
F = 1 - \Lambda r^2 - \frac{M}{r^{d-3}}.  
\end{equation}
Thus the two solutions perfectly agree in the large $r$ limit. It should however be noted that for the former the presence of $\Lambda$ is essential for this limit to exist. \\ 
  
Now we go to the general case and we write $G_{ab}^{(n)}$ for the differential operator resulting from the $n$-th term in the Lovelock polynomial and in particular, $G_{ab}^{(1)}$ is the Einstein tensor and $G_{ab}^{(2)} = H_{ab}$ of the Gauss-Bonnet. Again $(G^t_t = G^r_r)^{(n)}$ will give $\nu = -\lambda$, and  the analogue of Eq (12) would read as
\begin{equation}
(r^{d-2n-1} f^n)^\prime = \frac{2\Lambda}{d-2}r^{d-2}
\end{equation}
and it would readily integrate to give 
\begin{equation}
f^n = \Lambda_1 r^{2n} + \frac{k}{r^{d-2n-1}}.
\end{equation}
Again asymptotically it approximates to 
\begin{equation} 
F =1-\Lambda_1^{1/n} r^2-\frac{K}{r^{d-3}}
\end{equation}
where now $K=k/{n\Lambda_1^{1/n}}$. This is the $d$-dimensional Schwarzschild-dS/AdS black hole (for even $n$ only dS while for odd $n$ it could be dS/AdS with the sign of $\Lambda$). The corresponding Einstein-Lovelock solution is simply obtained by summing over $n$, $\sum_n \alpha_n $ on the left in Eq (17) with a coupling coefficient for each $n$. The polynomial on the left becomes difficult to solve for $n\ge4$. It is however expected that asymptotically the solution should tend to the Einstein solution in $d$ dimensions. We have seen that above for the quadratic $n=2$ Gauss-Bonnet case and we have also verified it for the cubic $n=3$ case  \cite{dehghani}.\\

Thus we establish that asymptotically the pure Lovelock for a given order $n$ and the Einstein-Lovelock ($\sum_n G_{ab}^n$) tend to the same limit, the Einstein solution in $d$-dimensional dS/AdS spacetime \cite{vac}. This is because higher order Lovelock contributions should wean out asymptotically, however what is rather interesting is the fact that even the order $n$ of the Lovelock polynomial does not matter so long as $\Lambda$ is  included. That is, the large $r$ limit is free of $n$ indicating a universal asymptotic behaviour.

\section{Uniform density sphere}

We know that the gravitational potential at any point inside a fluid sphere is given by $-M(r)/r^{d-3}$ for $d\geq4$ dimensional spacetime. Now $M(r)= \int \rho r^{d-2}dr$ which for constant density will go as $\rho r^{d-1}$ and then the potential will go as $\rho r^{d-1}/r^{d-3} = \rho r^2$ and is therefore independent of the dimension. This is an interesting general result: for the uniform density sphere, gravity has the universal character that it is independent of the dimension of spacetime and it always goes as $r$. It is then a natural question to ask, does this result carry over to Einsteinian gravity? In general relativistic language it is equivalent to asking, does the Schwarzschild interior solution always describe the field inside a uniform density sphere in all dimensions $d\geq4$? Not only that does it carry over to higher order Einstein-Gauss-Bonnet (Lovelock) gravity? The answer is yes \cite{unif}. \\

\subsection{Einstein case}
We begin with the general static spherically symmetric metric given by 
\begin{equation}
ds^2= e^\nu dt^2 - e^\lambda dr^2 - r^2d\Omega_{n-2}^2
\end{equation}
where $d\Omega_{n-2}^2$ is the metric on a unit $(n-2)$-sphere (in this Sec. $n$ refers to the dimension of spacetime). For the Einstein equation in the natural units ($8\pi G=c=1$),
\begin{equation}
G_{ab} = R_{ab} - \frac{1}{2} R g_{ab} = - T_{ab}
\end{equation}
and for perfect fluid, $T_a^b = diag(\rho, -p, -p, ..., -p)$,
we write 
\begin{equation}
e^{-\lambda}(\frac{\lambda^{\prime}}{r} - \frac{n-3}{r^2}) + \frac{n-3}{r^2} = \frac{2}{n-2}\rho  \label{density}
\end{equation}
\begin{equation}
e^{-\lambda}(\frac{\nu^{\prime}}{r} + \frac{n-3}{r^2}) - \frac{n-3}{r^2} = \frac{2}{n-2}p 
\end{equation}
and the pressure isotropy is given by 
\begin{eqnarray}
e^{-\lambda}(2\nu^{\prime\prime} + \nu^{\prime^2} - \lambda^{\prime}\nu^{\prime} - 2\frac{\nu^{\prime}}{r}) \nonumber \\
- 2(n-3)(\frac{e^{-\lambda}\lambda^{\prime}}{r} 
+ 2\frac{e^{-\lambda}}{r^2} - \frac{2}{r^2}) = 0  \label{isotropy}. 
\end{eqnarray}
We write this equation in a form that readily yields the universal character of the Schwarzschild interior solution for all $n\geq4$,  
\begin{eqnarray} 
e^{-\lambda}(2\nu^{\prime\prime} + \nu^{\prime^2} - \lambda^{\prime}\nu^{\prime} - 2\frac{\nu^{\prime} + \lambda^{\prime}}{r} - \frac{4}{r^2}) + \frac{4}{r^2} \nonumber \\ 
- 2(n-4) \Bigl((n-1)(\frac{e^{-\lambda}}{r^2} - \frac{1}{r^2}) + \frac{2\rho}{n-2} \Bigr) = 0. \label{iso}
\end{eqnarray}
We now set the coefficient of $(n-4)$ to zero so that it remains valid for all $n\geq4$. This then straightway determines $e^{-\lambda}$ without integration and it is given by 
\begin{equation}
e^{-\lambda} = 1 - \rho_0 r^2 \label{sol} 
\end{equation}
where $\rho_0=2\rho/{(n-1)(n-2)}$. This when put in Eq. (\ref{density}) implies constant density. We thus obtain $\rho=const.$ as the neceessary condition for universality of the isotropy equation for all $n\geq4$. The sufficient condition is obvious from the integration of Eq. (\ref{density}) for $\rho=const$, giving the same solution as above where a constant of integration is set to zero for regularity at the center. Thus constant density is the  necessary and sufficient condition for universality. Alternatively we can say that universality identifies uniform density. The universality is therefore true if and only if density is constant.\\

As is well known, Eq. (\ref{iso}) on substituting Eq. (\ref{sol}) admits the general solution as given by  
\begin{equation}
e^{\nu/2} = A + Be^{-\lambda/2} \label{sol2}
\end{equation}
where $A$ and $B$ are constants of integration to be determined by matching to the exterior solution. This is the Schwarzschild interior solution for a constant density sphere that is independent of the dimension except for a redefinition of the constant density as $\rho_0$. This proves the universality of the Schwarzschild interior solution for all $n\geq4$.\\

The Newtonian result that gravity inside a uniform density sphere is independent of spacetime dimension is thus carried over to general relativity as well despite nonlinearity of the equations. That is, Schwarzschild interior solution is valid for all $n\geq4$. Since there exist more general actions like Lovelock polynomial and $f(R)$ than the linear Einstein-Hilbert, it would be interesting to see whether this result would carry through there as well. That is what we take up next.  

\subsection{Gauss-Bonnet(Lovelock) case} 

The gravitational equation will now read as 
\begin{equation}
G^a_b +\alpha H^a_b = - T^a_b, \label{beq}
\end{equation}
where $H_{ab}$ as given in Eq (7). Now density and pressure would read as follows:
\begin{eqnarray}
\rho = \frac{(n-2)e^{-\lambda}}{2r^2} \Bigl(r\lambda^\prime - 
(n-3)(1-e^{\lambda})\Bigr)+\qquad\qquad\nonumber \\ 
+\frac{(n-2) e^{-2\lambda}\tilde\alpha}{2r^4}(1-e^{\lambda}) 
\Bigl(-2r\lambda^\prime + (n-5)(1-e^{\lambda})\Bigr) 
\label{rho-gb} 
\end{eqnarray}
\begin{eqnarray}
p = \frac{(n-2)e^{-\lambda}}{2r^2} \Bigl(r\nu^\prime + (n-3)(1-e^{\lambda})
\Bigr)-\qquad\qquad \nonumber \\
- \frac{(n-2) e^{-2\lambda}\tilde\alpha}{2r^4}(1-e^{\lambda}) \Bigl(2r\nu^\prime 
+ (n-5)(1-e^{\lambda})\Bigr)
\end{eqnarray}

The analogue of the isotropy Eq. (\ref{iso}) takes the form 
\begin{equation}
I_{GB} \equiv \biggl(1+\frac{2\tilde\alpha f}{r^2}\biggr){I_E}+
\frac{2\tilde\alpha}{r}\biggl(\frac{f}{r^2}\biggr)^{\prime}
\biggl[r\psi^\prime + \frac{f}{1-f}\psi\biggr] = 0 
\label{iso-gb}
\end{equation}
where $\psi=e^{\nu/2}, e^{-\lambda}=1-f, \tilde\alpha=(n-3)(n-4)\alpha$ and $I_E$ is given by LHS of Eq. (\ref{isotropy}), 
\begin{eqnarray}
I_E \equiv \frac{(1-f)}{\psi}\biggl\{\psi^{\prime\prime} -
\biggl(\frac{f^{\prime}}{2(1-f)}+\frac{1}{r}\biggr)\psi^{\prime} - 
\qquad\qquad\nonumber  \\
- \frac{(n-3)}{2r^2(1-f)}(rf^{\prime} - 2f)\psi\biggr\}. \label{isop}
\end{eqnarray}

From Eq. (\ref{rho-gb}), we write 
\begin{equation} 
(\tilde\alpha r^{n-5}f^2 + r^{n-3}f)^{\prime} = \frac
{2}{n-2}\rho r^{n-2}
\end{equation}
which integrates for $\rho=const.$ to give  
\begin{equation}
\tilde\alpha r^{n-5}f^2 + r^{n-3}f = \rho_0r^{n-1} + k	
\end{equation}
where $k$ is a constant of integration that should be set to zero for regularity at the center and $2\rho/(n-1)(n-2)=\rho_0$ as defined earlier. Solving for $f$, we get 
\begin{equation}
e^{-\lambda} = 1 - f = 1 - \rho_{0GB}r^2 \label{sol-gb}	
\end{equation}
where 
\begin{equation} 
\rho_{0GB} = \frac{\sqrt{1 + 4\tilde\alpha\rho_0} - 1}{2\tilde\alpha}.
\end{equation}
So the solution is the same as in the Einstein case and the  appropriate choice of sign is made so as to admit the limit $\alpha\rightarrow 0$ yielding the Einstein $\rho_0$ (the other choice would imply $\rho_{0GB}<0$ for positive $\alpha$). This, when substituted in the pressure isotropy Eq (\ref{iso-gb}), would lead to $I_E=0$ in Eq (\ref{isop}) yielding the solution (\ref{sol2}) as before. This establishes sufficient condition for universality. 

For the necessary condition, we have from Eq (\ref{iso-gb}) that either 
\begin{equation}
\Bigl(\frac{f}{r^2}\Bigr)^\prime = 0
\end{equation} 
or 
\begin{equation}
 r\psi^\prime + \frac{f}{1-f}\psi = 0.
\end{equation}

The former straightway leads with the use of Eq (33) to the same constant density solution (\ref{sol-gb}) and $I_E=0$ integrates to Eq. (\ref{sol2}) as before. This shows that universality implies constant density as the necessary condition. For the latter case, when Eq (38) is substituted in Eq (\ref{isop}) and $I_E = 0$ is now solved for $\lambda$, we again obtain the same solution (\ref{sol-gb}). Eq (33) again implies $\rho=const$ as the necessary condition. Now $\psi$ is determined by Eq. (38), which means the constant $A$ in solution (\ref{sol2}) must vanish. Then the  solution turns into de Sitter spacetime with $\rho=-p=const.$ which is a particular case of the Schwarzschild solution. This is, however, not a bounded finite distribution.\\

Thus universality and finiteness of a fluid sphere uniquely identifies the Schwarzschild interior solution for Einstein as well as Einstein-Gauss-Bonnet gravity. That is, gravity inside a fluid sphere of finite radius always goes as $r$ in all dimensions $\geq4$ so long as the density is constant. It is described by the Schwarzschild interior solution. It is only the constant density that gets redefined in terms of $\rho_0$ and $\rho_{0GB}$.\\

Though we have established the universality for EGB, it would in general be true for the  Einstein-Lovelock gravity. For that we argue as follows. The entire analysis is based on the two Eqs (\ref{iso-gb}) and (33). The GB contributions in the former appear in the multiplying factor to the Einstein differential operator, $I_E$ and another term with the factor $\tilde\alpha (f/r^2)^\prime$. This indicates that the contributions of higher orders in Lovelock polynomial will obey this pattern to respect quasilinearity of the equation. The higher orders will simply mean inclusion of the corresponding couplings in the multiplying factor as well as in the second term appropriately while the crucial entities, $I_E$ and $(f/r^2)^\prime$ on which the proof of the universality of the Schwarzschild solution hinges remain unchanged. On the other hand, Eq (33) is quadratic in $f$ for the quadratic GB action, which indicates that the degree of $f$ is tied to the order of the Lovelock polynomial. It essentially indicates that as $\rho_{0GB}$ is obtained from a quadratic algebraic relation, similarly in higher order its analogue will be determined by the higher degree algebraic relation. However the solution will always be given by Eq. (35). Thus what we have shown above  explicitly for EGB will go through for the general Einstein-Lovelock gravity. Thus the Newtonian result of universality of gravity inside a uniform density fluid sphere critically hinges on quasilinearity and would thus only be true for the Lovelock generalization of the Einstein gravity and not for any other. It could in a sense be thought of as yet another identifying feature of the Einstein-Lovelock gravity.\\ 

There is also an intriguing and unusual feature of GB(Lovelock) gravity. What happens if the multiplying  factor, $1+2\tilde\alpha f/r^2 = 0$ in Eq. (\ref{iso-gb})? Then the entire equation becomes vacuous, leaving $\psi$ completely free and undetermined while $e^{-\lambda} = 1 + r^2/2\tilde\alpha$. This leads to $p = -\rho = const. =  (n-1)(n-2)/8\tilde\alpha$, which is an anti-de Sitter distribution for $\alpha\ge0$. This is a special prescription where density is given by GB coupling $\alpha$. There is no way to determine $\psi$, and so we have a case of genuine indeterminacy of the metric. It is because GB(Lovelock) contributes such a multiplying factor involving ($\alpha,~ r,~ f$) to the second order quasilinear operator, which could be set to zero and thereby annulling  the equation altogether. This is a typical feature of Lovelock solutions which have been termed as geometrically free \cite{whitt}, \cite{whee}. Such a situation also occurs in the Kaluza-Klein split-up of six-dimensional spacetime into the usual $M^4$ and $2$-space of constant curvature in EGB theory ~\cite{mada}. \\

\section{Discussion}

Since gravity is universal it is always interesting to probe its universal features in various settings. We had set out to establish three universal features of gravity in higher dimensions. The first one was driven by the fact that gravitational equation of motion should always follow from the geometric properties of Riemann curvature even when higher orders in curvature are included. By suitably defining the analogue of Riemann as a homogeneous polynomial in Riemann, we have derived the analogue of the Einstein tensor by the vanishing trace of its Bianchi derivative for the any order in Lovelock polynomial. This is a novel characterization of the Lovelock polynomial \cite{bia}. Since it is derived from a geometric property, it is universal.\\

Next feature is driven by the physical consideration that higher order Lovelock effects  should appear as high energy corrections to the Einstein gravity and hence they should evaporate out in the low energy asymptotic limit, $r\to\infty$. That is all the vacuum solutions asymptotically tend to the Einstein solution in general $d$ dimensions. This is true for both pure Lovelock as well as Einstein-GB(Lovelock) solutions. However for the former non-zero $\Lambda$ is necessary for the asymptotic expansion. Here the universality  means that the asymptotic limit is independent of the Lovelock order \cite{vac}. \\

Finally it is obvious in the Newtonian theory that gravity inside a uniform density sphere is independent of the spacetime dimension simply because the dimensional change in both mass enclosed in certain radius and gravitational potential square out the spacetime dimension and hence it always goes as $r$. This is a general result that should be universally true for any good theory of gravity. We establish it for the Einstein as well as for Einstein-Lovelock gravity and its uniqueness. That is it would not be true for other generalizations like $f(R)$ theories. This is yet another characterization of the Lovelock gravity. It turns out that uniform density is necessary and sufficient condition for universality of the Schwarzschild interior solution \cite{unif}.

\acknowledgments

It is a pleasure to thank the organizers of the 13th Regional Conference on Mathematical Physics held on Oct. 23-31, 2010 at Antalaya, Turkey, where this was presented as a plenary lecture for the wonderful hospitality and exciting meeting.\\

\end{document}